\documentclass[11pt]{article}
\usepackage[margin=1in]{geometry}
\usepackage{graphicx}
\usepackage{hyperref}
\usepackage{booktabs}
\usepackage{amsmath}

\title{Salesforce Messaging Architecture: Platform Events, Async Sends, and Multi-Tenancy at Scale}
\author{Devam Gupta \\ Staff Engineer, Twilio \\ Technical Architect, Twilio for Salesforce \thanks{An earlier version of this paper was first posted on SSRN, Abstract ID 6903158: \url{https://papers.ssrn.com/sol3/papers.cfm?abstract_id=6903158}.}}
\date{July 2026}

\begin{document}
\maketitle

\begin{abstract}
Most enterprise messaging integrations function as external connectors. They reside outside the Customer Relationship Management (CRM) platform, receive webhooks, write peripheral logs, and consider the process integrated. Consequently, messages are isolated within the messaging platform while CRM records remain strictly inside the CRM, bridged only by fragile sync jobs, rigid field mappings, and eventual consistency windows. This paper outlines an alternative architectural paradigm: constructing the entire messaging core natively inside the CRM. Within this model, every message is treated as a native CRM record, every outbound path is a platform-native transaction, and all state parameters -- including delivery status, opt-in metrics, conversation history, and record ownership -- are stored as standard database objects. This structural native alignment allows the system to utilize standard reporting pipelines, operate inside native workflow builders, and trigger platform automation seamlessly. Drawing from a production managed package deployed across independent enterprise organizations spanning healthcare, sales operations, field services, and customer support, this study evaluates the core design patterns, platform scaling mechanics, multi-tenant decoupling constraints, and the boundaries of the platform-native architectural design.
\end{abstract}

\noindent\textbf{Keywords:} CRM Architecture, Enterprise Messaging, Salesforce, Webhooks, Platform Events, Data Modeling, Multi-Tenancy.

\section{Introduction and Related Work}

Enterprise messaging integrations have historically followed a connector model: the messaging platform remains the system of record, and the CRM receives a mirrored, delayed, and often incomplete copy of message state. This pattern is a direct descendant of the broker and message-router patterns catalogued by Hohpe and Woolf~\cite{hohpe2003}, which assume two independent systems bridged by asynchronous channels. That assumption is reasonable when the two systems have genuinely independent lifecycles -- but it becomes a liability when one of the two systems (the CRM) is also the primary interface through which humans reason about the customer relationship the messages belong to.

This paper documents an alternative to the connector model, evaluated against a production managed package (Twilio for Salesforce) distributed as a Salesforce first-generation managed package~\cite{sfisv2024} to independent enterprise organizations spanning healthcare, sales operations, field services, and customer support. Rather than treating the CRM as a downstream consumer of messaging events, the architecture treats the CRM as the sole system of record, and the external messaging gateway as a transport layer only. This inverts the conventional integration boundary and shifts the engineering burden from synchronization correctness to platform-native constraint management: transaction governor limits~\cite{sfgovlimits2024}, asynchronous execution ordering~\cite{sfqueueable2024}, and dynamic field- and object-level permission enforcement~\cite{sfperms2024} across tenants with heterogeneous configurations.

The remainder of this paper is organized as follows. Section 2 states the core architectural bet. Section 3 describes the data model. Sections 4 and 5 cover the outbound send path and the platform-event-based inbound path, including polymorphic record resolution at scale. Section 6 covers opt-in state as a data modeling decision, and is treated in substantially greater depth in a companion paper on consent architecture~\cite{gupta2026consent}. Section 7 covers status reconciliation, a topic expanded further in a forthcoming companion paper on webhook reliability~\cite{gupta2026webhook}. Section 8 covers multi-tenant constraints. Section 9 summarizes the architectural decisions, and Section 10 concludes.

\section{The Core Architectural Bet}

The fundamental design question governing any enterprise CRM integration is a simple one: \emph{where does the message record fundamentally live?}

In a traditional connector model, the system of record remains the messaging platform, while instances are mirrored down into the CRM. This approach introduces structural latencies: every report, analytic dashboard, or automated workflow involving message data must cross an API boundary or depend on a synchronization loop that operates behind real-time events.

Conversely, a CRM-native model establishes the CRM itself as the absolute system of record. The external messaging API is treated purely as a transport layer -- responsible for executing the delivery and returning immediate telemetry. The CRM is the system of record for everything else.

While conceptually straightforward, executing this pattern at enterprise scale introduces significant complexity when deployed across independent enterprise customer organizations -- each possessing unique object schemas, custom sharing rules, and volatile automation workloads.

\section{The Data Model}

The native architecture relies on a highly decoupled relational data model composed of six core objects:

\begin{itemize}
\item \textbf{Message Record (\texttt{Message\_\_c}):} Represents individual SMS units (inbound or outbound). It implements a polymorphic lookup schema to anchor to core CRM tables -- such as Contact, Lead, or Account -- along with a generic parent ID field to dynamically bind to custom enterprise objects.
\item \textbf{Campaign Send Record (\texttt{Campaign\_Send\_\_c}):} Represents high-volume batch dispatch operations and serves as the structural parent for grouped campaign messages.
\item \textbf{Opt-In Record (\texttt{Opt\_In\_\_c}):} Tracks individual opt-in and opt-out lifecycles per phone number per keyword. It is deliberately isolated, omitting hard foreign keys to individual Contacts to protect audit trails from record mutation or deletion.
\item \textbf{Configuration Setting:} A hierarchical custom metadata store containing secure API credentials mapped across global org-wide, profile-specific, and individual user-level scopes.
\item \textbf{Service Registry:} A routing index that maps external messaging platform identifiers (e.g., Messaging Service SIDs) to internal functional business units.
\item \textbf{Team / Team Membership:} Defines internal shared inbox groups, mapping specific CRM users to designated messaging transport lines.
\end{itemize}

The message record serves as the engine of this schema. An explicit metadata field (\texttt{Direction\_\_c}, bounded as \emph{inbound} or \emph{outbound}) controls UI thread rendering. By executing a simple query sorted chronologically and partitioned by this direction field, conversational threads are compiled dynamically on the client side without needing any external state machine or middle-tier indexing.

The external messaging API is treated purely as a transport layer -- responsible for executing the delivery and returning immediate telemetry.

\section{The Outbound Send Path}

Outbound message pipelines enforce a strict two-stage sequence: \textbf{record instantiation first, asynchronous network transmission second.}

This pattern is strictly mandated by the underlying CRM database engine. To prevent long-lived open transactions from locking database pools, synchronous execution blocks prohibit mixing Data Manipulation Language (DML) writes and external HTTP callouts within the same execution thread. The trigger writes the outbound record to disk synchronously with a state of \emph{null} or \emph{Queued}. A secondary asynchronous thread (Queueable~\cite{sfqueueable2024}) is instantly spun up after the parent transaction commits, assuming the network overhead, firing the payload to the external gateway, and recording the returned transaction ID and delivery telemetry.

This design ensures the message record exists on disk before the network layer is engaged. If a severe network failure, gateway timeout, or remote outage occurs, the local audit record remains intact -- allowing operations to query, trace, and execute automated retries with full visibility.

\subsection{Managing Scale under Transactional Governor Limits}

Enterprise systems cap the maximum number of distinct HTTP callouts allowed within a single asynchronous transaction thread at exactly 100~\cite{sfgovlimits2024}. Consequently, a bulk marketing campaign or emergency alert dispatching 5,000 messages cannot compile linearly inside a singular job context.

To overcome this constraint, the system utilizes a programmatic self-chaining recursive loop. The executing class tracks the precise number of callouts completed. As the transaction count nears 95, the job halts linear execution, slices the remaining allocation array, instantiates a new instance of itself with the residual data packet, and enqueues the child job into the platform's queue manager.

Furthermore, high-volume multi-tenant environments introduce a secondary concurrency failure mode: the platform's global asynchronous queue can become saturated by competing enterprise workflows, causing immediate enqueue rejections. To preserve systemic reliability, the outbound architecture embeds a fallback scheduler. If the system encounters a queue capacity rejection, it catches the platform exception, initializes a micro-timer, and schedules the remaining payload to fire via an independent cron job 20 seconds later. Messages are safely deferred, ensuring zero-drop guarantees under high load.

\section{The Inbound Path: Webhook Decoupling via Platform Events}

Inbound data pipelines face different operational hazards, necessitating a reverse structural pattern.

The public gateway interface consists of an unauthenticated REST custom controller registered directly as the destination webhook URL on the telecommunications platform. When an inbound text arrives from a mobile network, the gateway executes an HTTP POST against this CRM endpoint.

A standard linear implementation would process the inbound payload inside the body of the web service handler, executing record searches, evaluating routing rules, inserting database entities, and generating push notifications before responding to the network socket. This pattern introduces severe risk at scale: complex CRM business logic execution can easily drift past 2,000 to 3,000 milliseconds, exceeding the gateway's strict webhook timeout thresholds. This causes the gateway to treat the request as a dropped connection, dropping the transaction and triggering immediate retry sequences that flood the database with duplicate entries and race conditions.

To decouple network latency from internal execution overhead, the architecture places an enterprise event bus directly behind the web service layer using native Platform Events~\cite{sfplatformevents2024}.

The incoming REST thread acts solely as a high-speed validation proxy. It confirms the cryptographic signature (HMAC), validates the account identifiers against the metadata cache, publishes a lightweight event containing the raw payload string to the platform event bus, and instantly returns an HTTP code 202 Accepted back to the gateway. The network socket is released in less than 50 milliseconds. The actual record processing, data matching, and relational linkage execute in a separate, isolated transaction driven by the platform event consumer thread, completely insulated from the external gateway's timeout clocks.

\section{Polymorphic CRM Record Linking and Bulk Resolution}

When an inbound message event is picked up by the asynchronous worker, the payload presents only a raw E.164 phone string. The application must accurately resolve this string to a specific Contact, Lead, Account, or custom business entity. Because platform event triggers consume records in chunks of up to 200 items per execution block, performing sequential database queries per message is impossible -- this would trigger an immediate $N+1$ query exception and crash the inbound thread.

The system handles this through a strict set-based bulk map resolution pattern:

\begin{enumerate}
\item All incoming phone strings across the entire execution array are scrubbed, normalized, and collected into a single unique index array.
\item The engine executes exactly one optimized database search query per targeted entity type (Contact, Lead, Account) matching against the full aggregated string array.
\item The resulting records are mapped into an in-memory runtime index table (\texttt{Map<String, Id>}).
\item The worker iterates through the message batch, checks the index maps in sequence, and populates the matching polymorphic lookup reference field.
\end{enumerate}

To extend this to proprietary custom entities without altering the underlying package codebase, the architecture exposes an injection metadata registry. This registry allows system administrators to declare external custom tables and target phone fields. The runtime engine consumes this metadata file and uses dynamic Structured Object Query Language (SOQL) to append the custom fields to the bulk query pipeline, achieving out-of-the-box support for specialized sectors like healthcare patient records or recruitment workflows.

\section{Opt-In State as a Data Model Decision}

A common flaw in many messaging integrations is reducing opt-in management to a simple boolean checkbox flag (e.g., \texttt{Opted\_In\_\_c = True}) directly on the Contact or Lead record. This approach introduces structural anti-patterns that quickly break down in enterprise environments:

\begin{itemize}
\item \textbf{Granularity failure:} An individual person record frequently contains multiple phone numbers (e.g., home, work, mobile). A binary attribute on the parent record cannot track opt-in variations across distinct device endpoints.
\item \textbf{Loss of audit compliance:} Binary fields inherently overwrite their own history. They lack temporal resolution, making it impossible to audit the exact chronological point where opt-in was granted, transferred, or rescinded.
\item \textbf{Query performance degradation:} Filtering an outbound list of 500,000 marketing targets against fluid conditions on parent tables requires complex table joins that degrade database optimizer performance at scale.
\end{itemize}

The architecture resolves this by defining an opt-in record as an independent, standalone opt-in object (\texttt{Opt\_In\_\_c}) keyed structurally to a phone number and a localized tracking keyword, completely decoupled from the contact table's relational lifecycle.

The opt-in record uses an explicit composite key constructed from the normalized E.164 phone string and the tracking keyword identifier (e.g., \texttt{+15550199\_MARKETING\_ID}). When incoming opt-out command flags are intercepted (e.g., STOP, QUIT), the engine executes an atomic database upsert operation directly against this composite key index. This ensures absolute idempotency: even if network failures cause the same opt-out payload to be processed multiple times, the transaction updates the exact same record without creating duplicates or risking data corruption.

A full treatment of this consent data model -- including keyword lifecycle management, bulk consent lookup across large recipient lists, and TCPA-relevant audit requirements -- is given in a companion paper~\cite{gupta2026consent}.

\section{Systemic Resilience: The Two-Path Status Verification Model}

In an enterprise architecture, assuming network-level operations will always execute perfectly is a major vulnerability. Status updates sent via webhooks face multiple real-world failure modes: target platforms go offline for scheduled maintenance windows, network routes drop packets under load, and high-volume batch campaigns can generate rapid webhook callbacks that trigger rate limiters on the entry controllers.

To guarantee eventual data correctness across all records, the system implements a distinct Two-Path Status Verification Model.

\begin{itemize}
\item \textbf{Path 1 (The Reactive Real-Time Pipeline):} Operates via the high-speed webhook platform event architecture described in Section 5. It handles immediate status changes (queued $\rightarrow$ sent $\rightarrow$ delivered) in real time, serving as the primary delivery path.
\item \textbf{Path 2 (The Proactive Reconciler):} An independent cron daemon running via scheduled platform classes at regular intervals. This background job requests transactional logs directly from the gateway API for the preceding window, isolates records whose local states have drifted from the remote status, and forces reconciliation updates.
\end{itemize}

Because both pipelines target the exact same immutable external gateway transaction identifier, the processing framework remains completely idempotent. If Path 1 fails due to an outage, Path 2 gracefully captures the update on its next cycle. If both paths succeed, the redundant write resolves cleanly with zero side effects. This fallback approach guarantees absolute state accuracy even under severe structural disruptions.

The failure modes, reconciliation cadence, and idempotency guarantees of this two-path model are analyzed in greater depth in a forthcoming companion paper on webhook reliability~\cite{gupta2026webhook}.

\section{Multi-Tenant Architecture Challenges}

Distributing a uniform codebase as a managed package across independent enterprise customer environments introduces architectural complexities that are completely invisible when developing for a single isolated organization.

\subsection{Enforcing Security and Access Controls Dynamically}

The runtime framework cannot make hard assumptions about the data access controls or object-level security permissions configured in any given installation. In one environment, the executing user might operate under global administrative clearance; in another, strict Field-Level Security (FLS) may block visibility to the primary phone fields.

To prevent unauthorized data access or severe runtime faults, the application isolates all DML operations behind a centralized security abstraction proxy layer. Raw database commands like \texttt{insert} or \texttt{update} are strictly banned across the codebase. Instead, all transactions pass through a unified access class that evaluates runtime user tokens, inspects target table access parameters, validates field permissions~\cite{sfperms2024}, and either executes the safe database operation or surfaces a safe handled exception.

\subsection{Handling Object Trigger Conflicts}

When the core application creates an inbound contact or registers a message, it operates inside the same database transaction pool as the customer's proprietary local workflows. If an enterprise client has deployed resource-heavy triggers or long-running validation rules on their Contact table, those scripts consume execution time that counts against the shared transaction limits.

To mitigate transaction resource exhaustion, the package exposes an architecture bypass framework. It provides a public, state-preserving memory class that local developers can check within their custom code blocks (e.g., \texttt{if (Package.isMessagingActive()) return;}). This allows clients to intelligently suspend non-critical local processes during high-speed messaging transactions, preventing limit crashes.

\subsection{The Operational Cost of Schema Permanence}

Under managed package distribution rules~\cite{sfisv2024}, database schema modifications are strictly additive. Once an object table or data field is published to production, it can never be deleted from the package manifest. This permanence makes early data modeling choices a long-term operational tax.

For example, the generic parent ID string field currently used to support custom entities is a direct result of this constraint. It stands as an architectural scar from a legacy design phase that predated the platform's modern polymorphic lookup features. Because removing it would break existing consumer databases, it must be maintained indefinitely, requiring future iterations to be engineered alongside legacy structural paths.

\section{Summary of Key Architectural Decisions}

Table~\ref{tab:summary} summarizes the structural core and scalability parameters of the native architecture.

\begin{table}[h]
\centering
\caption{Key Architectural Decisions Summary Matrix}
\label{tab:summary}
\begin{tabular}{p{5.5cm}p{7.5cm}}
\toprule
\textbf{Architectural Decision Pattern} & \textbf{Core Functional Guarantee \& Mitigation Benefit} \\
\midrule
Message record inserted before API send & Guarantees audit trail regardless of send outcome. \\
Platform Event for inbound webhook & Decouples webhook response latency from record processing. \\
Self-chaining async job for bulk sends & Respects callout governor limits without hard failures. \\
Fallback scheduler on job queue limit & Defers sends rather than dropping them under platform pressure. \\
Opt-in state as a separate object with composite key & Per-number granularity, full history, idempotent bulk operations. \\
Two-path status sync & Webhook failures degrade gracefully; polling catches missed updates. \\
Centralized DML framework layer & All CRUD/FLS enforcement in one place; security vulnerabilities fixed at a single checkpoint. \\
Generic parent ID field & Retrofitted polymorphic custom support without breaking schema paths. \\
\bottomrule
\end{tabular}
\end{table}

\section{Conclusion and Future Horizons}

The native CRM-native design model offers clear operational advantages for enterprise workflows: it eliminates complex external ETL pipelines, enables zero-code automation via native visual tools, utilizes standard access control layers, and provides out-of-the-box compatibility with next-generation autonomous AI orchestration platforms like Agentforce.

However, achieving these benefits requires a deliberate engineering trade-off: developers must build around rigid transaction governor limits, handle complex asynchronous debugging paths, and design schemas with absolute permanence in mind. By utilizing systematic patterns -- such as platform event webhooks, recursive self-chaining jobs, and independent ledger-style opt-in models -- engineers can build high-volume, reliable communication networks that scale seamlessly inside enterprise environments.

Future work includes extending the governed-action pattern described here to autonomous AI agent platforms (e.g., Salesforce Agentforce), where the same centralized DML/FLS enforcement layer and multi-tenant isolation guarantees must be preserved when actions are triggered by an LLM-driven agent rather than a human user or a deterministic trigger.

\end{document}